\documentclass[aps,prb,reprint,showpacs,showkeys,preprintnumbers,groupedaddress,amsmath,amssymb,]{revtex4-1}

\usepackage{graphicx}
\usepackage{dcolumn}
\usepackage{bm}
\usepackage{hyperref}
\usepackage[dvips]{color}

\begin{document}




\title{Universal Hall coefficient correction in strongly coupled Cr-SiO$_2$ nanogranular metals}


\author{Bin Zheng}
\affiliation{Tianjin Key Laboratory of Low Dimensional Materials Physics and
Preparing Technology, Department of Physics, Tianjin University, Tianjin 300354,
China}
\author{Zhi-Hao He}
\affiliation{Tianjin Key Laboratory of Low Dimensional Materials Physics and
Preparing Technology, Department of Physics, Tianjin University, Tianjin 300354,
China}
\author{Zhi-Qing Li}
\email[Author to whom correspondence should be addressed. Electronic address: ]{zhiqingli@tju.edu.cn}
\affiliation{Tianjin Key Laboratory of Low Dimensional Materials Physics and
Preparing Technology, Department of Physics, Tianjin University, Tianjin 300354,
China}


\date{\today}

\begin{abstract}
The microstructure and electrical transport of Cr$_x$(SiO$_2$)$_{1-x}$ nanogranular films with Cr volume faction $x$$\simeq$0.67 and 0.72 are systematically investigated. The transmission electron microscopy images and elemental mappings indicate that the films are quite inhomogeneous: some
Cr granules directly connect to others while some Cr granules with size $\sim$1 to $\sim$3\,nm disperse in the SiO$_2$ dielectric matrix. For each film, the Hall coefficient $R_H$ varies linearly with the natural logarithm of temperature, i.e., $\Delta R_H$$\propto$$\ln T$, above $\sim$60\,K, saturates at $\sim$60\,K, and retains the saturating value below $\sim$60\,K. The temperature dependence of Hall coefficient can be explained by the recent theory in granular metals and originates from \emph{virtual diffusion} of electrons through the metallic granules. For the conductivity $\sigma$, a robust $\Delta \sigma$$\propto$$\sqrt{T}$ law is observed from $\sim$50 down to 2\,K. The behavior of the conductivity stems from the ``Altshuler-Aronov" correction, whose influence on the Hall coefficient is not present in the films.
\end{abstract}


\maketitle


Granular metals are a new class of artificial functional materials, in which the metal granules are embedded in insulating matrix (usually amorphous). The ease of adjusting granule size and the ratio of the metal to insulator makes granular metal a perfect model system for the investigation of the interplay of electronic correlations, quantum confinement effects, and disorder. It has been realized for quite some time that granular metal can reveal new physical phenomena which is absent in homogeneous disordered systems due to its specific nanoscale structure.\cite{Beloborodov-2007,X. X. Zhang-2001,Y. N. Wu-2010} The early extensive investigations on the properties of the granular metals can be dated back to a century ago,\cite{Garnett1904,Garnett1906,Swann-PM-1914} and the earlier advances had been reviewed in Refs.~\onlinecite{Abeles-1975,Abeles-1976}. Recently significant progress has been made in this field.\cite{Beloborodov-2007,Efetov-2003,Beloborodov-2003,Kharitonov-2007,Kharitonov-2008,Mulligan-2016,Feigelman-2018,Fan-2018,Wu-2018,Gnanasekaran-2019} As for the electrical transport properties, one of the new important theoretical findings\cite{Kharitonov-2007,Kharitonov-2008} is the influence of the quantum effects of the Coulomb interaction on Hall coefficient in strong intergrain coupling limit. According to Kharitonov and Efetov,\cite{Kharitonov-2007,Kharitonov-2008} the tunneling escape energy\cite{Escape-energy} $g_T\bar{\delta}$ (with $g_T$ being the intergrain tunneling conductance in the unit $2e^2/\hbar$ and $\bar{\delta}$ the mean energy level spacing in a grain) is an important watershed with respect to the electrical transport properties of granular metals. In a wide temperature range $T$$\gtrsim$$T^\ast$ ($T^\ast$$\equiv$$ g_T\bar{\delta}/k_B$), quantum effects of the Coulomb interaction lead to a logarithmic in $T$ correction to the Hall coefficient, which is specific to granular metals and absent in homogeneous disordered conductors. The specific correction saturates at $T^\ast$ and keeps as a constant below $T^\ast$. The ``Altshuler-Aronov'' correction to the longitudinal conductivity may govern the temperature behavior of Hall coefficient at much lower temperatures ($T$$\ll$$ T^\ast$). It should be emphasized the predicted $\ln T$ behavior of Hall coefficient is valid both in two dimensional (2D) and three dimensional (3D) granular metals and thus is universal.

In 2D granular metals, the theoretical prediction of Kharitonov and Efetov\cite{Kharitonov-2007,Kharitonov-2008} has been verified experimentally in ultrathin Sn doped In$_2$O$_3$ films\cite{Y. J. Zhang-2011} and ultrathin Al doped ZnO films.\cite{Y. Yang-2012} However, the $\ln T$ behavior of Hall coefficient in 3D granular metals has not been reported for more systems other than  Ag-SnO$_2$ granular films.\cite{Y. N. Wu-2015} In addition, the temperature behavior of Hall coefficient below $T^\ast$ has not been experimentally explored thus far. Therefore, it is necessary to completely investigate the Hall transport properties of 3D granular metals at temperatures below and above $T^\ast$. To measure the temperature dependence of Hall coefficient below $T^\ast$, one should enhance the tunneling escape energy $g_T\bar{\delta}$. Since the mean energy level spacing $\bar{\delta}$ is proportional to $1/V$ ($\bar{\delta}$$=$$1/\nu V$ with $\nu$ being the density of states at the Fermi energy and $V$ the volume of the grain), an effective way to increase the tunneling escape energy is to reduce the mean grain-size $a$. Considering the metal with low melting point easily forms large size grains in granular composites, we need choose metals with relative high (at least moderate) melting points and high resistance to oxidation as the metallic constituent. We notice that chromium metal not only possesses relative high melting point (2176\,K) and high resistance to oxidation at room temperature, but also has relative low carrier concentration.\cite{Ziman,Hsieh-1996,Kittel,Ashcroft} The low carrier concentration character facilitates the measurement of the temperature dependence of Hall coefficient. In addition, chromium metal is immiscible with SiO$_2$.\cite{Abeles-1976} Thus Cr-SiO$_2$ granular films may be good candidates to test the temperature dependence of Hall transport properties below and above $T^\ast$. In the present paper, we have systematically investigated the microstructure and Hall transport properties as well as the longitudinal conductivities of Cr$_x$(SiO$_2$)$_{1-x}$ (the Cr volume fraction $x$$\simeq$0.67 and 0.72). The influence of the quantum effects of the Coulomb interaction on Hall coefficient is fully addressed.

Cr$_x$(SiO$_2$)$_{1-x}$ films with different metal volume fraction $x$ ($0.5$$\lesssim $$x$$\lesssim $$1$) were deposited at room temperature by co-sputtering method. A Cr and a SiO$_2$ targets both with 99.99\% in purity were used as the sputtering source. The base pressure of the chamber is less than $1\times 10^{-4}$\,Pa. The metal volume fraction $x$ was controlled via adjusting the sputtering powers applied in the two targets. The films were simultaneously deposited on the glass (Fisherfinest premium microscope slides) and polyimide (Kapton) substrates for the transport and composition measurements, respectively. Hall bar shaped films, defined by using mechanical masks, were deposited for Hall coefficient and resistivity measurements. Cr/Au electrodes were fabricated to obtain good contact.

The thicknesses of the films, ranging from 600 to 700\,nm, were measured using a surface profiler (Dektak, 6 M). The Cr volume fraction of the sample was determined by energy-dispersive x-ray spectroscopy analysis (EDS, EDAX, Model Apollo X) using the film deposited on Kapton. The microstructure of the films was characterized by transmission electron microscopy (TEM, Tecnai G2 F20 S-Twin, operating at 200\,kV). The micro-area distribution of Cr, Si, and O elements were scanned by EDS (FEI Super X) both in cross-sectional and in-plane TEM samples. The electrical conductivity and Hall coefficient were measured using a physical property measurement system (PPMS-6000, Quantum Design) by employing the standard four-probe method. In the Hall coefficient measurement process, the magnetic field was set to scan from $-4$ to $4$\,T in a step of 0.2\,T at a certain testing temperature. In the case of conductivity measurements, a magnetic field of 7\,T perpendicular to the film plane was applied to suppress the weak-localization
effect.\cite{Lin-2001}

\begin{figure}
\begin{center}
\includegraphics[scale=1]{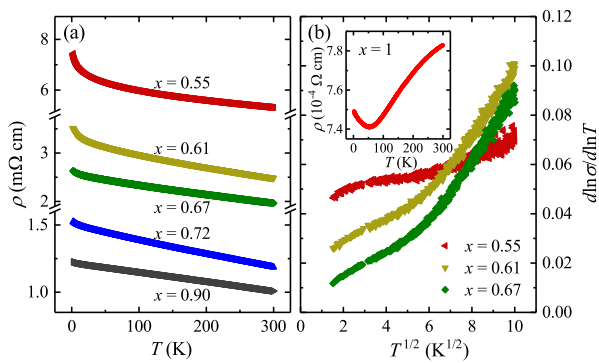}
\caption{(Color online) (a) Resistivity versus temperature for some representative films. (b) $\mathrm{d}$$\ln$$\sigma/\mathrm{d}$$\ln $$T$ versus $T^{1/2}$ for the $x$$\simeq$0.55, 0.61, and 0.67 films. Inset: resistivity versus temperature for pure Cr film. }\label{FigR-x}
\end{center}
\end{figure}

Figure~\ref{FigR-x}(a) shows the normalized resistivity as a function of temperature for some representative films as indicated. For the pure Cr film, the resistivity decreases with decreasing temperature from 300 down to $\sim$50\,K, reaches its minimum at $\sim$50\,K, and increases with further decreasing temperature [Inset of Fig.~\ref{FigR-x}(b)]. This is the typical characteristics for disordered metals.\cite{Kittel,Lin-2001,Lee-1985} While for the Cr$_x$(SiO$_2$)$_{1-x}$ films, the resistivities increase with decreasing temperature in the whole measured temperature range even if the volume fraction $x$ reaches $\sim$0.9. Superficially, the characteristic of $\rho$-$T$ curves predicts these Cr$_x$(SiO$_2$)$_{1-x}$ ($x$$\lesssim$0.9) films are insulator in electrical transport properties. However, from 300 down to 2\,K, the changes of the resistivities are less than 40\% even for the $x$$\simeq$0.55 film.  We analyze the temperature behavior of the logarithmic
derivative of the conductivity $w$$=$$\mathrm{d}$$\ln$$\sigma/\mathrm{d}$$\ln$$ T$,\cite{A. Mobius-2003,M.Huth-2011} which represents a more sensitive method to reliably decide whether a certain sample is metallic or insulating.
Figure~\ref{FigR-x}(b) shows $w$ versus $T^{1/2}$ for the $x$$\simeq$$0.55$, 0.61, and 0.67 films. As the temperature approaches to zero, the values of $w$ for the $x$$\simeq$0.55 and 0.61 films tend to a constant, while $w$ tends to zero for the $x$$\simeq$0.67 film. The nearly zero value of $w$ means a metallic film while a constant or infinite values of $w$ imply a insulating film.  Hence the percolation threshold, below which the films are insulating (or semiconducting), lies between 0.61 and 0.67. Considering the results obtained from the 0.67$\lesssim$$x$$\lesssim$0.90 are similar, we only present and discuss the results obtained from two representative metallic films ($x$$\simeq$$0.67$ and 0.72) below.

\begin{figure}
\begin{center}
\includegraphics[scale=0.9]{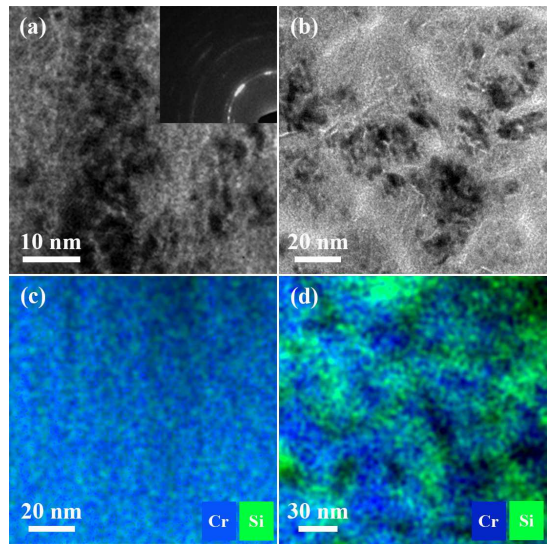}
\caption{(Color online)  Cross-sectional (a), and in-plane (b) TEM images for the $x$$\simeq$0.67 film. Elements (Cr and Si) distribution  mappings in cross-sectional (c), and in-plane (d) TEM samples. Inset in (a): electron diffraction patterns for the film.}\label{FigTEM}
\end{center}
\end{figure}

Figure~\ref{FigTEM}(a) and \ref{FigTEM}(b) show the bright-field cross-sectional and in-plane TEM images for the $x$$\simeq$0.67 film, respectively. From the figures, one can see that the distribution of Cr granules is quite inhomogeneous: some Cr granules (the dark regions) directly connect to others while some Cr granules with size $\sim$1 to $\sim$3\,nm disperse in the SiO$_2$ dielectric matrix (the bright regions). Only weak diffraction rings related to (110) plane of the body-centered cubic Cr can be observed in the electron diffraction patterns [inset of Fig.~\ref{FigTEM}(a)], indicating that the SiO$_2$ is amorphous. The micro-area element analysis is carried out to obtain the detailed distribution information of Cr and SiO$_2$. Figure~\ref{FigTEM}(c) and \ref{FigTEM}(d) show the elemental distribution of Cr and Si in cross-section and in-plane area of the $x$$\simeq$0.67 film, respectively. The distribution map of oxygen element (not shown) is almost identical to that of Si, indicating that the oxygen combines with Si instead of Cr. Close inspection of Fig.~\ref{FigTEM}(c) and \ref{FigTEM}(d) indicates that lots of Cr clusters with dimensions $\sim$1 to $\sim$3\,nm are randomly dispersed in the Si (i.e., SiO$_2$) matrix. This is in accordance with the result of the TEM images.

\begin{table}
\caption{\label{Table1} Relevant parameters for the films. Here $x$ is Cr volume fraction, $t$ is the mean film thickness, $n^\ast$ is effective carrier concentration, and $g_T$ is the intergrain tunneling conductance in unit $2e^2/\hbar$. $g_T\bar{\delta}/k_B$ is the characteristic temperature defined in the text, and $\min(g_T E_c, E_{\rm Th})/k_B$ is the upper bound temperature for Eq.~(\ref{Eq.(Hall)}) to hold.}
\begin{ruledtabular}
\begin{center}
\begin{tabular}{ccccccc}
$x$        &$t$    & $n^\ast$               &$g_T$            &   $g_T\bar{\delta}/k_B$  & $g_TE_c/k_B $ &  $E_{\mathrm{Th}}/k_B$   \\
     & (nm)  & ($10^{28}\,$m$^{-3}$)    &   &         (K)              & (K)           &   (K)           \\ \hline

0.67    &663    & 5.61                   &0.21        &          42        & 390           &1520       \\
0.72    &643    & 6.48                   &0.22        &          45        & 408            &1520       \\
\end{tabular}
\end{center}
\end{ruledtabular}
\end{table}

\begin{figure}
\begin{center}
\includegraphics[scale=1]{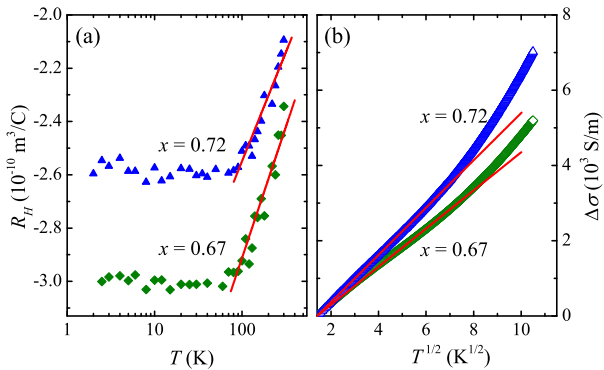}
\caption{(Color online) (a) The Hall coefficient versus temperature for the $x$$\simeq$0.67 and 0.72 films. The solid straight lines are least-square fits to Eq.~(\ref{Eq.(Hall)}). (b) Conductivity change versus $T^{1/2}$ for the two films. The solid straight lines are least-square fits to Eq.~(\ref{Eq-Cond}).} \label{FIGRH-T}
\end{center}
\end{figure}

Figure~\ref{FIGRH-T}(a) shows the variation of Hall coefficient $R_H$ with logarithm of temperature for the $x$$\simeq$0.67 and 0.72 films, as indicated. The Hall coefficients of the Cr$_x$(SiO$_2$)$_{1-x}$ granular films are negative, which is different to that of the pure Cr film.\cite{Hsieh-1996,C. S. Lu-1967,E. Osquiguil-2013} From Fig.~\ref{FIGRH-T}(a), one can see that the Hall coefficient of each film varies linearly with $\log T$ ($\ln T$) from 300 down to $\sim$80\,K, and becomes temperature independent below $\sim$60\,K. The temperature dependent behavior of Hall coefficient is similar to that predicted by Kharitonov and Efetov.\cite{Kharitonov-2007,Kharitonov-2008} Next we quantitatively compare the $R_H(T)$ data with the theoretical prediction.

According to Kharitonov and Efetov,\cite{Kharitonov-2007,Kharitonov-2008} when the quantum effects of the Coulomb interaction and weak localization are considered, the Hall coefficient of the granular metal can be written as,
\begin{equation}\label{Eq.(Hall)}
R_H=\frac{1}{n^\ast e} \left[ 1+\frac{c_{d}}{4\pi g_T} \ln \left(\frac{E_0}{\max(k_{B}T,g_T\bar{\delta})}\right)-2\frac{\delta\sigma^{AA}(T)}{\sigma_0} \right],
\end{equation}
where $n^\ast$ is the effective carrier concentration, $c_d$ is a numerical lattice factor of order unity, $E_0$$=$$\min(g_{T}E_{c}, E_{\text{Th}})$ is the upper bound energy for the validity of Eq.~(\ref{Eq.(Hall)}), and $\sigma_0$ is the longitudinal conductivity without the influence of Coulomb interaction and interference effects. The first term in the square brackets is the classical Hall coefficient of the system ($R_H^0$$=$$1/n^\ast e$), the second term is the correction of \emph{virtual diffusion} (VD) of electrons through the metallic granule, and the third term represents ``Altshuler-Aronov'' correction to longitudinal conductivity. The weak localization correction to the Hall coefficient is zero. The VD correction term, which is specific to granular metals and absent in homogeneous disordered metals, saturates at $T$$\simeq$$T^\ast$ and retains the saturating value below $T^\ast$.  The ``Altshuler-Aronov'' correction to longitudinal conductivity vanishes at $T$$>$$T^\ast$, while at $T$$<$$T^\ast$ and for 3D granular metal it can be expressed as\cite{Beloborodov-2007,Beloborodov-2003}
\begin{equation}\label{Eq-Cond}
\frac{\delta\sigma^{AA}}{\sigma_0}=\frac{1.83}{12\pi^2g_T}\sqrt{\frac{k_BT}{g_T\bar{\delta}}}, \hspace{3 mm} d=3.
\end{equation}
Eq.~(\ref{Eq-Cond}) is similar to that derived from homogeneous disordered conductors.\cite{Alsthuler-1985,Lee-1985,Altshuler-1980,Altshuler-1980-1}

Theoretically, Eq.~(\ref{Eq.(Hall)}) is valid at $g_T$$\gg$$1$. Recently, it has demonstrated that the validity condition of Eq.~(\ref{Eq.(Hall)}) can be expanded to $g_T$$>$$g_T^c$,\cite{Y. N. Wu-2015,M.Huth-2011} where $g_T^c$=$(1/2\pi d)\ln(E_c/\bar{\delta})$ (with $d$ being the dimensionality) is the critical tunneling conductance,\cite{Beloborodov-2003} above which the system lies in the metallic regime.
Treating Cr granules as roughly spherical and taking the mean-diameter $a$$\simeq$2\,nm, one can readily obtain the theoretical value of $T^\ast$ is $\sim$45\,K.\cite{free-electron}  We compare the experimental $R_H(T)$ data above 60\,K with the prediction of Eq.~(\ref{Eq.(Hall)}). The solid straight lines in Fig.~\ref{FIGRH-T} represent the least-square fits results. In the fitting processes,  the third term in the square brackets is neglected, $n^{\ast}$ and $g_T$ are adjustable parameters, the lattice factor $c_d$ is set as $c_d$$=$$1$, the charging energy is $E_c$$=$$e^2/(4\pi\epsilon_0\epsilon_r a)$ with dielectric constant $\epsilon_r$$=$$4.5$ for SiO$_2$,\cite{K. F. Young-1973} and the Thouless energy is $E_{\mathrm{Th}}$$=$$\hbar D/a^2$ (where $D$ is the electron
diffusion constant). The fitted values of $n^{\ast}$ and $g_T$, as well as some related parameters for the two films, are summarized in Table~\ref{Table1}. For the 3D Cr$_x$(SiO$_2$)$_{1-x}$ films, the critical tunneling conductance is estimated to be $g_T^c$$\simeq$0.12, which is less than the $g_T$ values of $x$$\simeq $$0.67$ and 0.72 films. The theoretical upper bound temperature $\mathrm{min}(g_T E_c, E_{\mathrm{Th}})/k_B$ for Eq.~(\ref{Eq.(Hall)}) holding is $\sim$400\,K. Thus the valid condition of Eq.~(\ref{Eq.(Hall)}) is fully fulfilled and the ln$T$ behavior of the Hall coefficient originates from the VD of electrons inside individual grains.

In fact, Eq.~(\ref{Eq.(Hall)}) predicts $\delta R_H/R_H^0$$=$$-2\delta \sigma^{AA}/\sigma_0$ at $T$$<$$T^\ast$. To analyze the low temperature behavior of the Hall coefficient, we first discuss the temperature dependence of conductivity at $T$$<$$T^\ast$. In granular metals, it has been found that the corrections of the quantum effect of Coulomb interaction to the conductivity are composed of high energy and low energy contributions with respect to the characteristic energy $g_T\bar{\delta}$.\cite{Beloborodov-2007,Beloborodov-2003,Efetov-2003} The high energy correction is proportional to $\ln T$ at $T$$>$$T^\ast$, saturates at $T$$=$$T^\ast$, and keeps as a constant at $T$$<$$T^\ast$. The low energy correction can be ignored at $T$$>$$T^\ast$, and has been expressed as Eq.~(\ref{Eq-Cond}) at $T$$<$$T^\ast$ in 3D systems. Thus in 3D systems the change of the conductivity $\Delta\sigma(T)=\sigma(T)-\sigma(T_0)$ can be written as \emph{} $\Delta\sigma(T)=\delta\sigma^{AA}(T)-\delta\sigma^{AA}(T_0)$ at $T$$<$$T^\ast$, where $T_0$ is an arbitrary reference temperature.
Figure~\ref{FIGRH-T}(b) shows $\Delta\sigma$ ($T_0$$=$2\,K) as a function of $T^{1/2}$ under 7\,T for the $x$$\simeq$0.67 and 0.72 films (the field is high enough to suppress the weak localization effect\cite{Lin-2001}). Clearly, the $\Delta\sigma$ data vary linearly with $T^{1/2}$ below $\sim$50\,K. The solid straight lines in Fig.~\ref{FIGRH-T}(b) are the least-square fits to Eq.~(\ref{Eq-Cond}), in which $g_T$ is the adjustable parameter and $\sigma_0$ is  approximately taken the conductivity at 100\,K. For the $x$$\simeq$0.67 and 0.72 films, the fitted values of $g_T$ are 0.19 and 0.24, respectively. Thus the $g_T$ values obtained independently from fitting Eqs.~(\ref{Eq.(Hall)}) and (\ref{Eq-Cond}) are almost identical for each film, which not only verifies Eq.~(\ref{Eq-Cond}) but also confirms the validity of the Eq.~(\ref{Eq.(Hall)}) at $T$$>$$T^\ast$. Since the contribution of the electron-phonon scattering to $\sigma$ (but not to $R_H$) becomes progressively notable at high temperature regime, the theoretical predicted logarithmic temperature behavior of $\sigma$ is not observed in high temperature regime.

We return again to discuss the Hall coefficient of the films. The nearly temperature independent feature of the Hall coefficients below $\sim$60\,K  [Fig.~\ref{FIGRH-T}(a)] indicates the theoretical predicted $\delta R_H(T)$$\propto$$\sqrt{T}$ behavior [Eq.~(\ref{Eq.(Hall)}) and Eq.~(\ref{Eq-Cond})] at  $T$$<$$T^\ast$ is not present in the 3D Cr$_x$(SiO$_2$)$_{1-x}$ granular films. Thus the constant Hall coefficient below $\sim$60\,K come from VD correction term besides $R_H^0$ [Eq.~(\ref{Eq.(Hall)})]. In fact, similar theoretical result, $\delta R_H/R_H^0$$=$$-2\delta \sigma^{AA}/\sigma_0$, has also been derived in homogeneous disordered conductors.\cite{Lee-1985,Altshuler-1980-1,Altshuler-1980,Alsthuler-1985} This prediction has been experimentally verified in different homogeneous disordered 2D conductors.\cite{Bishop-1980,Uren-1980,Bergmann-1984,Zhang-2015} However the experimental results in homogeneous disordered 3D conductors are quite inconsistent. For example, it has been experimental found that the Hall coefficients of the CuTi and Pd$_{80}$Si$_{20}$ alloys\cite{A. Schulte-1984,Howson} is independent of temperature, while the change of the Hall coefficients are proportional to $\sqrt{T}$ for the Pd$_{30}$Zr$_{70}$, Ni$_{64}$Zr$_{36}$, Ni$_{24}$Zr$_{76}$, and Fe$_{24}$Zr$_{76}$ alloys.\cite{A. Schulte-1984,B.L. Gallagher-1984} Thus the temperature dependence of Hall coefficient of granular metals at $T$$<$$T^\ast$ need further experimental investigations.

In summary, we have investigated the Hall transport properties and longitudinal conductivities of 3D Cr$_x$(SiO$_2$)$_{1-x}$ nanogranular films lying in the metallic regime. It is found that the films have relative high tunneling escape energy and the temperature behaviors of Hall coefficient are dominated by the correction of VD of electrons inside individual grains over the whole measured temperature range. The  ``Altshuler-Aronov'' correction to the conductivity governs the temperature dependence of conductivity at $T$$<$$T^\ast$. However, the theoretical predicted relation,  $\delta R_H/R_H^0$$=$$-2\delta \sigma^{AA}/\sigma_0$, is not observed. Our experimental results fully demonstrate the validity of the theoretical predictions concerning the unique VD correction to Hall coefficient in granular metals.

This work is supported by the National Natural Science Foundation of China through Grant No. 11774253.


\begin{thebibliography}{00}
\bibitem{Escape-energy} We let $\tau_0$ be the time that the electron spends inside a grain, then the tunneling escape rate from one grain to the nearest neighbors is $\tau_0^{-1}$. Considering the relation $g_T\bar{\delta}$$=$$\hbar \tau_0^{-1}$, thus we designate the energy $g_T\bar{\delta}$ as tunneling escape energy.
\bibitem{free-electron} Free-electron model is used in calculating the value of $\bar{\delta}$ and $D$. Considerding the Hall coefficient of the Cr film is positive, we take the carrier concentration of $x$$\simeq$0.90 film in calculating processes.
\end{thebibliography}
\end{document}